\documentstyle[12pt]{article}

\newcommand{\la}[1]{\label{#1}}

\newlength{\numlen}
\newcommand{\n}{\settowidth{\numlen}{0}\makebox[\numlen]{}}
\newcommand{\cen}[1]{\multicolumn{1}{c}{#1}}

\newlength{\indexlength}

\newcommand{\be}{\begin{equation}}
\newcommand{\ee}{\end{equation}}
\newcommand{\ba}{\begin{eqnarray}}
\newcommand{\ea}{\end{eqnarray}}
\newcommand{\eps}{\varepsilon} 
\newcommand{\rmi}[1]{{\mbox{\scriptsize #1}}}

\newcommand{\etal}{{et al.\ }}
\newcommand{\eq}{eq.~}
\newcommand{\eqs}{eqs.~}
\newcommand{\fig}{fig.~}

\newcommand{\nr}[1]{(\ref{#1})}
\newcommand{\h}{{\hspace{0.5 cm}}}

\newcommand{\dd}{\mbox{d}}

\newcommand{\latent}{L}
\newcommand{\oabs}{\Omega_{\rmi{abs}}}
\newcommand{\orot}{\Omega_{\rmi{rot}}}

\def\SigRotF{0.0292(22)}
\def\SigRotS{0.0218(33)}
\def\SigAbsF{0.0295(21)}
\def\SigAbsS{0.0218(33)}

\begin{document}

\begin{titlepage}
\mbox{}\hfill UTHEP-252\\
\mbox{}\hfill AZPH-TH/93-04\\
\mbox{}\hfill CERN-TH.6798/93
\begin{centering}
\vfill

{\bf INTERFACE TENSION IN QUENCHED QCD}

\vspace{1cm}
Y. Iwasaki$^a$, K. Kanaya$^a$, Leo K\"arkk\"ainen$^b$,\\
K. Rummukainen$^c$, and T. Yoshi\'e$^a$

\vspace{1cm}
{\em $^a$Institute of Physics, University of Tsukuba,\\
Ibaraki 305, Japan\\}

\vspace{0.3cm}
{\em $^b$Department of Physics, University of Arizona,\\
Tucson, AZ 85721, USA\\}

\vspace{0.3cm}
{\em $^c$Theory Division, CERN,\\ CH-1211 Geneva 23, Switzerland}

\vspace{2cm}
{\bf Abstract}

\end{centering}

\vspace{0.3cm}\noindent
We calculate the tension $\sigma$ of the interface between the confined
and deconfined phases by the histogram method
in SU(3) lattice gauge theory for temporal
extents of 4 and 6 using the recent high-statistics data by QCDPAX
collaboration.  The results are $\sigma/ T_c^3 = \SigRotF$ and
\SigRotS\ for $N_t=4$ and 6, respectively.
The ratio $\sigma/ T_c^3$ shows
a scaling violation similar to that
already observed for the latent heat $\latent$.  However,
we find that the physically interesting dimensionless combinations
$(\sigma^{3}/\latent^2 T)^{1/2}$ and $\sigma T/\latent$  scale
within the statistical errors.

\vfill \vfill
\noindent
UTHEP-252\\
AZPH-TH/93-04\\
CERN-TH.6798/93\\
September 1993
\end{titlepage}

\section{Introduction}

At the transition temperature of a first-order phase transition, 
a mixed state can exist where two different bulk phases are
separated by an interface.  The free
energy densities of the two bulk phases are equal and the free energy of
the mixed state is higher than either of the pure phases by the amount
of $F_s = \sigma A$, where $A$ is the area of the interface and
$\sigma$ is the interface tension.  
If the high temperature phase transition of QCD is of first order,
the interface tension between
hadronic matter and quark-gluon plasma is an important parameter for the
formation of quark-gluon plasma in heavy ion
collisions and for the nucleation of hadronic matter in the early Universe;
in the latter case the inhomogeneities generated during the phase
transition could affect the light element
nucleosynthesis~\cite{Cosmo}.

In recent years many methods have been introduced to 
compute numerically
the interface tension in lattice SU(3) gauge theory.  Because
the measurement of the interface tension generally requires an
extensive computing effort, so far only lattices with
small temporal extents (values
of $N_t$) have been used.  Most of the work has been concentrated on
the $N_t=2$ system~\cite{nt2papers,Grossmann92,Berg92}, some reaching
$N_t=4$~\cite{Potvin91,Grossmann93}.

In this work we measure the interface tension in SU(3) gauge theory
with the {\em histogram
method\,} introduced by K.~Binder~\cite{Binder82}.  It requires
high-statistics histograms of an order parameter at or near the phase
transition point.  We analyze the Polyakov loop histograms obtained
by QCDPAX collaboration~\cite{QCDPAX92}
with the recent high-statistics simulations on $N_t=4$ and $N_t=6$ lattices.

Dynamics of the first order phase transition is largely governed by
the latent heat $\latent$ and the interface tension $\sigma$ -- the
``driving'' and ``braking'' forces of the transition.  
It was already observed by QCDPAX collaboration that
the dimensionless quantity $\latent/T_c^4$ does {\em not} scale,
when $N_t$ is increased from 4 to 6. 
As reported below, we observe
similar behavior for the interface tension $\sigma$.  This makes it
very difficult to estimate the continuum limit values of both $\latent$ and
$\sigma$. However, when one studies physical
processes occurring during the phase transition, $\sigma$ and
$\latent$ appear in certain combinations.  We find that the scaling
properties of these combinations are much improved.  

Let us briefly consider, as an example, the initial stage of the first 
order phase transition in the early Universe. 
Let us assume that the system initially exists in the 
high temperature
phase and is cooled below $T_c$ with a (constant) rate $C = -\partial
T/\partial t$.  According to the classical nucleation theory
\cite{Landau80}, near $T_c$ the system spontaneously nucleates small
bubbles of the low temperature phase.  The nucleation rate is
$\Gamma\,\dd t\dd V \propto T^4 e^{-F_B/T} \dd t\dd V$, where
$F_B=-\latent\hat T\, V_B+\sigma A_B$ is the free energy of the bubble
of volume $V_B$ and area $A_B$, and $\hat T = (T_c-T)/T_c$.  When
$T<T_c$, $F_B$ has a maximum when the radius is
$R_c=2\sigma/(\latent\hat T)$; a bubble with this radius is called a
critical bubble.  The free energy of the critical bubble is given by
$F_c=F_B(R_c)=\alpha^2\hat T^{-2}\, T_c$, where
\be
  \alpha^2 = 16\pi\sigma^3/3\latent^2 T_c \,. 
  \la{alpha}
\ee
If a bubble larger than $R_c$ is nucleated, it starts to expand.
Expanding bubbles emit shock waves which reheat the supercooled matter
back close to the transition temperature, thus hindering the formation
of the new bubbles close to the old ones.  The shock waves propagate
approximately with the sound velocity $v_s$.  New bubbles can be
nucleated only in the fraction of the volume unaffected by the shock
waves. This volume fraction is~\cite{Cosmo}
\be
  f(t) = 1 - \int_{t_c}^{t} \dd t' \frac{4}{3}\pi v_s^3 (t-t')^3
  \,\Gamma[T(t')].
\ee
The bubble nucleation is finished at the time $t_\rmi{pt}$ when
$f(t_\rmi{pt})\approx 0$.  After some algebra, the amount of
supercooling and the average distance between the nucleated bubbles
turn out to be, 
\be
  T_c - T_\rmi{pt} \approx \alpha \chi^{-1/2}\,T_c\,,
  \la{bubble1}
\ee
and
\be
  d \approx   v_s\pi^{1/3} e^{\chi/4} \alpha \chi^{-3/2}\,T_c^{-1}\,,
  \la{bubble2}
\ee
respectively,
where $\chi = 4\log(T^2_c/C)$.  For the QCD phase transition in the
early Universe~\cite{Cosmo}, $\chi\approx 173$, $T_c\approx 150$\,MeV
and $v_s^2 \approx 1/3$. The equations above give $T_c-T_\rmi{pt}
\approx 0.076\alpha\,T_c$ and $d\approx 3\alpha\mbox{\,m}$.  The
interface tension and the latent heat appear only in the combination
$\alpha$ 
in this case.

In section \ref{hist-section}, we describe the histogram method:
we briefly introduce the histogram method and apply it to the $SU(3)$
gauge theory.
Then we consider finite size
corrections to get the interface tension in the infinite
volume limit.  The results are given in section \ref{results} and the
conclusions are given in section \ref{conclusions}.

\section{The Histogram Method}
\label{hist-section}

\subsection{The Probability Distribution}
\label{dist-section}

In this section we introduce the histogram method \cite{Binder82} and
apply it to the deconfinement phase transition of SU(3) gauge theory.
Let us consider SU(3) gauge theory on a lattice of the size $V\times
N_t\,a$, where $V = N_x\times N_y\times N_z\,a^3$, with $a$ being the
lattice spacing.  On an Euclidean lattice, the temperature is given by
$T = 1/(N_t\,a)$.  Close to the transition coupling $\beta_c$
($\beta=6/g^2$), the probability distribution $P(\Omega)$ of a
variable $\Omega$, that has a discontinuity at the thermodynamical
limit, develops a double-peak structure. For example, the order
parameter $\Omega$ can be the absolute value of the Polyakov loop.
The peaks correspond to the pure phase configurations and between the
peaks the dominant contributions come from mixed state configurations
where the phases are separated by interfaces.  When the volume is
increased, the peaks become more pronounced.  The suppression of the
mixed configurations is caused by the
extra free energy of the interfaces. 
With the labeling of the directions such that $N_x$ and $N_y \le N_z$,
the interfaces form preferentially in the ($x$-$y$) -plane.  For
definiteness, in what follows we assume that the interfaces are
oriented along the ($x$-$y$) -plane; the effects of the non-favored
orientations are discussed in the section \ref{fss-section}.  Because
of the periodic boundary conditions, there are two interfaces with a
total area $2A = 2\,N_x N_y\,a^2$.

We assume that, in large enough volumes, the
distributions around the peaks are described by Gaussians:
\be
  c_i\exp[-(\Omega-\Omega_i)^2/d_i^2]\,, \la{gauss}
\ee
where $i=1,2$ labels the two peaks, and $d_i\propto V^{-1/2}$,
$c_i\propto V^{1/2}$. 
Here $\Omega_1$ and $\Omega_2$ are 
the values of an extensive variable
in the confined and the deconfined
phases, respectively.
We assume that, for the time being, $\Omega$ is a real-valued variable. 
The complication which arises in the case of the complex Polyakov loop
will be discussed later.
The probabilities for the system to reside in the confined or deconfined
phases are given by, respectively, $\exp(-f_1 V/T)$ and $\exp(-f_2 V/T)$,
except for a common normalization factor.
Here $f_1$ and $f_2$ are the free energy densities of the confined 
and deconfined phases,
respectively, at the given $\beta$ and $N_t$:
the free energy densities of both phases can be defined by a partial partition
function summed over
a subset of configurations which belong to each Gaussian distribution.

We assume that in the mixed state between the
peaks the two phases occupy two parts of the volumes, $V_1$ and $V_2$, 
separated by two interfaces.  The
probability for the mixed phase is
\be
  P_m(\Omega) =  c_m
  \exp[-(f_1\,V_1+f_2\,V_2)/T - \sigma\,2A/T].
  \la{intprob}
\ee 
where $\sigma$ is the interface tension and $V_1+V_2=V$.
$P_m$ depends on $\Omega$ only through the relation
\be
  \Omega = (\Omega_1 V_1 + \Omega_2 V_2)/V \,.
  \la{OmegaV}
\ee
Then, the whole
probability distribution is given by
\be
  P(\Omega) = P_1(\Omega) + P_2(\Omega) + P_m(\Omega),
  \la{distribution}
\ee
with
\be
  P_i(\Omega)=c_i \exp(-f_i V/T) \exp[-(\Omega-\Omega_i)^2/d_i^2]\ \ \ (i=1,2)
  \la{disgaus} 
\ee
Equation~\nr{distribution}
is our basic assumption on the distribution of the variable.
The coefficient $c$'s in equations~\nr{intprob} and 
\nr{disgaus} 
depend on the power of volume $V$.

At the critical coupling $\beta_c$ in the infinite volume limit,
the free energy
densities $f_1$ and $f_2$ are equal.
On a finite lattice, however, in general $f_1$ and $f_2$ are not equal.
When $f_1$ and $f_2$ are equal, the distance between the two
interfaces can vary without changing the total free energy:
$P_m$ is a constant independent of $\Omega$. Then 
$P(\Omega)$ becomes a sum of two Gaussians and a constant.
On the other hand, when $f_1$ and $f_2$ are not equal,
there is no flat part
connecting the two peaks.

Let us denote the
two maxima of $P(\Omega)$ by $p_{\max,1}$ and $p_{\max,2}$ and the
minimum between the peaks by $p_{\min}$.  
Let the minimum point given by 
\be
 {d\over{d\Omega}}P(\Omega)=0
\ee
equal to 
\be
\Omega=\gamma_1\Omega_1+\gamma_2\Omega_2
\la{mini}
\ee
with $\gamma_1+\gamma_2=1$, which implies $V_1=\gamma_1 V, V_2=\gamma_2 V$.
Then, the leading volume dependence, that is, the $V_i$ dependence in the
exponent,
cancels in the quantity
\be 
\hat\sigma_{V} \equiv -\frac{N_t^2}{2N_x N_y}\,\log\frac{p_{\min}}
	{(p_{\max,1})^{\gamma_1}\,(p_{\max,2})^{\gamma_2}}.
\la{hatsigma}
\ee
which converges to the interface tension divided by $T_c^3$
when $V\rightarrow\infty$:
\be
\hat\sigma \equiv \sigma/T_c^3 =
	  \lim_{V\rightarrow\infty} \hat\sigma_{V}\,.
\la{tension}
\ee
Note that when $p_{\max,1}=p_{\max,2}=p_{\max}$,
the denominator in the logarithm 
${(p_{\max,1})^{\gamma_1}\,(p_{\max,2})^{\gamma_2}}$
reduces to $p_{\max}$.
The effect of the volume dependence of the coefficient $c$'s will
be taken into account,
when we consider finite size corrections in the next section.

Eq.~\nr{tension} gives the
interface tension for each $N_t$.
To get the continuum
value we still have to take the limit $N_t\rightarrow\infty$
($a\rightarrow 0$) while keeping $T=T_c$.

When one applies the histogram method, one has to do some decisions
which, in principle, are irrelevant in the infinite volume limit, but
can cause systematic differences when working at finite volumes.  Let
us first elaborate further on the choice of the variable
we adopt in the
analysis.  Previous studies of SU(3) gauge theory on $N_t=2$ lattices
have used the action density~\cite{Berg92} or the Polyakov
loop~\cite{Grossmann92,Grossmann93}; we chose the latter because 
only the Polyakov loop
gives a clear separation of phases for the all cases we have investigated.

The Polyakov loop
$\Omega$ is a complex-valued observable, and near $\beta_c$ the
distribution $p(\Omega)$ develops 4 peaks, corresponding to the
confined phase at the origin and 3 degenerate deconfined phases to the
directions $\exp(i2\pi n/3)$, $n=0,1,2$.  While, in principle, one
could apply \eq\nr{hatsigma} to a complex-valued order parameter, this
would require excessive statistical quality of the data.  We projected
$\Omega$ to the real axis by either using the absolute value 
($\oabs$) or by
rotating it ($\orot$), 
by multiplying it with $\exp(i2\pi n/3)$, to the sector
$-\pi/3 <\mbox{arg\,}\Omega \le \pi/3$ and taking the real part.

The projection $\Omega\rightarrow\oabs$ or $\orot$ causes a
deformation in the probability distribution near the origin. Let us
assume that, in a large enough volume, the original distribution of
the peak at the origin is given by $P(\Omega) = C
\exp(-|\Omega|^2/d^2)$, where $C\propto V$ and $d\propto V^{-1/2}$.
The projection changes this to
\begin{eqnarray}
  P(\oabs) &=& 
  C\,d\,2\pi(\oabs/d)\exp(-\oabs^2/d^2)  \\
  P(\orot) &=& 
  C\,d\,3\sqrt{\pi} \mbox{\,erf\,}(\sqrt{3}\orot/d)
  \exp(-\orot^2/d^2)
\end{eqnarray}
The maxima of these distributions are 
$\sqrt{2}\pi e^{-0.5} \times C\,d$ 
for $P(\oabs)$ and $3.2354\ldots\times C\,d$ for $P(\orot)$, where
$C\,d\propto V^{1/2}$, in accord with \eq\nr{gauss}.  On the other
hand, the modifications in the other peaks away from the origin should be
small and can be neglected. 

Now let us consider what happens for $\hat\sigma_V$ in these cases.
After some algebras we find that 
the $\hat\sigma_V$ in this case also is given by \eq\nr{hatsigma}
with minimum point \eq\nr{mini}.

The second choice concerns the value of $\beta$ used in the analysis.
Using the reweighting technique~\cite{reweighting} we adjusted $\beta$
so that the peaks of the histograms for $\orot$ had equal height.
Alternatively, one may use $\beta$ where the ratio of the 
weights of the peaks is 1/3, corresponding to 1 disordered and 3
ordered phases; for the 2-dimensional $q$-state Potts models, this
method has been shown to yield the correct infinite volume $\beta_c$
up to exponentially suppressed corrections~\cite{Borgs91}.  However,
we did not apply this procedure here, because for the smallest volumes
available to us, the minimum point cannot be clearly
identified to apply \eq\nr{hatsigma}.
Further, as discussed above, we can expect that
\eq\nr{hatsigma} is stable with respect to the changes in $\beta$.
Indeed, 
we have checked that the $\hat\sigma_V$'s calculated at $\beta_{\rm run}$
agree with those calculated at $\beta_0$
within the statistical errors.
The final values of $\beta_0$ used in the analysis are listed
in table \ref{table1}.  The actual distributions of
$P_{\beta_0}(\orot)$ are shown in \fig\ref{fighisto}, normalized so
that $p_{\max} = 1$.

Thirdly, we have to specify the way the heights are measured from the
histograms. Due to statistical noise one should not just use the
minima and maxima of the histograms in the analysis.  To suppress
random fluctuations we performed a third order polynomial fit 
to the histograms near the extrema and calculated the extremum values from
the fits.  The fit ranges were chosen by consulting the shape of the
histogram by eye; again, the results remained virtually unchanged,
when different ranges were used.

\subsection{Finite Size Corrections}

\label{fss-section}
In this section, we discuss various finite size corrections(FSC)
to \eq\nr{tension}.  Recently there have been a lot of
interests in the analysis of FSC to the interface tension
\cite{Wiese92,Bunk92,Caselle92}; the analysis is particularly
important here because of the different geometries of the available
lattices.  The starting assumptions of the analysis are that the
interfaces are infinitely thin and that they do not interact with each
other.  We take into account the following three 
possible finite size effects:
Gaussian fluctuations of the order parameter in the bulk
phases, Gaussian capillary wave fluctuations of the
interfaces and the zero mode of the translation of interfaces
to the direction
perpendicular to the interfaces.  Altogether, the ansatz becomes
\be
\frac{p_{\min}}{p_{\max}} = \sum_{i=1}^3
	Z_\rmi{bulk} Z_\rmi{zero} Z_\rmi{cw}^2
	e^{-2 N_j N_k \hat\sigma/N_t^2}\,,\h 
        \hat i \perp \hat j \perp \hat k
\la{fspartition}
\ee
where $i$ labels the direction perpendicular to the interfaces (of an
area $N_j\,N_k a^2$). Here $p_{\max}$ stands for 
${(p_{\max,1})^{\gamma_1}\,(p_{\max,2})^{\gamma_2}}$.
In the ansatz \nr{fspartition}, $\hat\sigma$ is assumed to be
independent of $N_x$, $N_y$, and $N_z$, but does in general depend on $N_t$
(if we are not in the scaling region).

The bulk fluctuation term
arises from the Gaussian fluctuations of the order parameter.  The
width of the Gaussian is proportional to $V^{-1/2}$, so that the
normalization is $\propto V^{1/2}$.  Because $p_{\max}$ is in the
denominator in \eq\nr{fspartition},
\be
Z_{\rmi{bulk}} \propto 1 / \sqrt{N_x N_y N_z}.
\ee

The zero-mode contribution which arises from the translations of the two
interfaces to direction $z$ with keeping the relative distance 
$\ell$ 
fixed is proportional to $N_z(\sqrt {N_x N_y})^2$ 
and the contribution from the infrared divergent part of 
the capillary wave fluctuation is proportional to $(1/\sqrt {N_x N_y})^2$.
Together they give a factor
\be
Z_\rmi{zero} \propto N_z^2\,,
\la{zeroc}
\ee
an extra power of $N_z$ originating from the measure:
${\rm d}\Omega = (\Omega_1 - \Omega_2) N_z^{-1} {\rm d}\ell$.

The capillary wave fluctuation term excluding the infrared divergent term 
given by Bunk~\cite{Bunk92} 
and Caselle \etal\cite{Caselle92} is:
\be
Z_{\rmi{cw}}(N_x,N_y) \propto
 \sqrt{\frac{N_x}{N_y}}\,[\eta(i\,N_y/N_x)]^{-2},
\la{zcw}
\ee
where $\eta$ is the Dedekind eta function:
\be
\eta(\phi) = e^{2\pi i \phi/24}
	\prod_{n=1}^\infty \,(1-e^{2\pi i n\phi}).
\la{eta}
\ee
Note that $Z_{\rmi{cw}}$ depends only on the ratio
$N_x/N_y$ of the interface.  Not so apparent is the fact that
$Z_{\rmi{cw}}(N_x,N_y) = Z_{\rmi{cw}}(N_y,N_x)$.
Because there are two interfaces, the $Z_{\rmi{cw}}$ term is squared
in \eq\nr{fspartition}.

When deriving \eqs\nr{zeroc} and \nr{zcw}, we have considered 
the interfaces 
in the $z$ direction only, while in \eq\nr{fspartition} we consider
all interfaces including those in the $x$ and $y$ directions.

In our lattices the shorter spatial lengths are equal: $N= N_x=N_y$.
Rewriting \eq\nr{fspartition}, we obtain the formula for 
$\hat\sigma_V$ defined by \eq\nr{hatsigma}:
\be
\hat\sigma_V(N,N_z,N_t) =
     \hat\sigma - \frac{N_t^2}{N^2} \left(
           c  + \frac{3}{4} \log {N_z} - \frac{1}{2}\log{N} 
              + \frac{1}{2}G(N,N_z,\hat\sigma) \right) \, ,
\label{solvsig}
\ee
where $c$ is a constant
independent of $N$ and $N_z$ 
and
\be
G(N,N_z,\hat\sigma) = 
\log \left(1 + 2\Bigr(\frac{N}{N_z}
\frac{Z_{\rmi{cw}}(N,N_z)}{Z_{\rmi{cw}}(N,N)}\Bigl)^2 
\exp[{-2(N_z-N)N \hat\sigma/N_t^2}] \right) .
\label{geometric}
\ee
The term $G$ appears because we have taken into account the interfaces
along ($x,z$)- and ($y,z$) -planes and the value of it crucially depends on the
geometric feature of the lattice.
For example, for cubic volumes $G=\log 3$ and for long cylinders
$G=0$; when the lattice is long ($N_z \gg N$), 
the effect of the planar interfaces in less advantageous
directions is  exponentially suppressed.
In our case, the effect of $G$ is significant only for
the more cubical $N_t=6$ lattices.  
Because the analytic value of $c$ is not so far available, we
perform a least-mean-squares fit of \eq\nr{solvsig} to the data with 
parameters $\hat\sigma$ and $c$ in the both $N_t$ cases.

Higher excitations, like multiple wall ($4,6,\ldots$ walls)
configurations or configurations containing a large spherical droplet
of one phase embedded in the other phase, can also contribute to the
probability density between the peaks.  However, they have considerably
larger interface area and are correspondingly more suppressed.
Therefore, we neglect their effects as well as the interaction
between the two interfaces. 

\section{Results}

\label{results}

\begin{table}
\center
\begin{tabular}{ccrlllll}\hline
  \cen{$N_t$} & \cen{$N_xN_yN_z$} & \cen{Iterations}& 
  \cen{$\beta_\rmi{run}$} & 
  \cen{$\beta_0$} &
  \cen{$\hat\sigma_{V}\,(\Omega_\rmi{rot})$}&
  \cen{$\hat\sigma_{V}\,(\Omega_\rmi{abs})$}&
  \cen{$\gamma_1(\oabs)$} \\ \hline
4 & $12^2\times 24$&   910,000 & 5.6915 & 5.69115 & 0.0282(28)&0.0241(27) & 0.580(37)\\
4 & $24^2\times 36$&   712,000 & 5.6925 & 5.69261 & 0.0308(17)&0.0300(16) & 0.556(21)\\
6 & $20^3 $        &   376,000 & 5.8922 & 5.89049 & 0.0146(34)&0.0123(28) & 0.569(101)\\
6 & $24^3 $        &   480,000 & 5.89   & 5.89197 & 0.0158(26)&0.0143(22) & 0.547(62)\\
6 & $36^2\times 48$& 1,112,000 & 5.8936 & 5.89402 & 0.0173(25)&0.0164(26) & 0.597(42)\\
\hline
\end{tabular}
\caption[1]{Parameters of the runs and the results for $\hat\sigma_V$.
  $\beta_\rmi{run}$ is the
  value used in the actual simulation, and $\beta_0$ is the reweighted
  value.  $\hat\sigma_V$ is calculated from \eq\nr{hatsigma}, using
  both the rotated and the absolute value of the Polyakov
  line histograms at $\beta_0$. 
  $\gamma_1$ is measured from the $\oabs$ histogram at $\beta_0$. 
  \la{table1}}
\end{table}

The parameters of the simulations are gathered in Table~\ref{table1}.
The simulations were performed with $\beta=\beta_\rmi{run}$.
This was
reweighted to $\beta_0$ in order to obtain ``equal height'' histograms
for $\orot$
(\fig\ref{fighisto}).  The values of $\hat\sigma_V$ [\eq\nr{hatsigma}]
for $\orot$ and $\oabs$, as well as the values of $\gamma_1$ 
for $\oabs$ at $\beta_0$, 
are also presented in Table~\ref{table1}. 
The error analysis was performed with
the jackknife method, with the bin sizes identical to those used
in ref.~\cite{QCDPAX92}.

Let us first give the results of 
the interface tension in the infinite volume limit obtained for the rotated
Polyakov line:
\be
\hat\sigma_\rmi{rot} = \sigma/T_c^3 = \left\{
	\begin{array}{ll}
	 \SigRotF & \h (N_t = 4) \\
	 \SigRotS & \h (N_t = 6)
	\end{array}\right. 
\la{sigmaresult}
\ee
The values of the constant $c$ in \eq\nr{solvsig} are $-1.307(43)$
and $-1.230(58)$ for $N_t=4$ and $6$, respectively.
For $N_t=6$, the $\chi^2$ values are quite acceptable 
($\chi^2/{\rm dof}=0.31$). This implies that 
our assumption for FSC is reasonable.

The results obtained for the absolute value of the Polyakov line 
are in remarkable agreement with
the values given in \eq\nr{sigmaresult}:
\be
\hat\sigma_\rmi{abs} = \left\{
	\begin{array}{ll}
	 \SigAbsF & \h (N_t = 4) \\
	 \SigAbsS & \h (N_t = 6)
	\end{array}\right.\,
\ee
(For $N_t=6$, the difference can be seen only in the fourth
significant digit).  The values of the constant $c$ are
$-1.267(41)$ ($N_t=4$) and $-1.206(53)$ ($N_t=6$), and 
$\chi^2/{\rm dof}=0.49$ for $N_t=6$.
We can conclude that the results do not depend on
the choice of $\orot$ or $\oabs$ for the order parameter.
For definiteness, we use only the rotated
values in the further analysis.

Let us look closer at the effects of the FSC and, in
particular, the function $G$.  For $N_t=4$, the linear fit to the data
gives $\hat\sigma=0.0317(24)$ without FSC, $0.0301(24)$ with
FSC but with $G=0$, and $0.0292(24)$ with the full FSC ansatz -- 
the first two values agree with the last value 
within the statistical errors. 
On the other hand, when $N_t=6$, the corresponding values
are $\hat\sigma=0.0185(39)$ (without FSC), 0.0330(39) ($G=0$), 
and 0.0219(39) 
(full FSC ansatz).
In
this case the function $G$ cancels most of the correction introduced
by the other terms in \eq\nr{solvsig}.

The $N_t=4$ value above is consistent with the value $\hat\sigma =
0.027(4)$ measured by Brower \etal~\cite{Potvin91} using a completely
different method.  Recently, Grossmann and Laursen \cite{Grossmann93}
performed a histogram method analysis for $N_t=4$ and 2
lattices, using the histograms of the $N_t=4$ lattices published in
refs.~\cite{QCDPAX92,Fukugita89}.
Their result \cite{Grossmann93} 
$\hat\sigma=0.025(4)$ is consistent with our result.
(Although the value 
$\hat\sigma=0.040(4)$ which originally appeared in \cite{Grossmann93}, 
is somewhat inconsistent with our value, they recalculated it.) 
For $N_t=2$, they obtained the value $\hat\sigma = 0.092(4)$.

The numbers above which are listed in Table~\ref{table2}
indicate that the ratio ${\sigma}/{T_c^3}$ does not
exhibit scaling behavior for $N_t=2$ -- 6, and not even for $N_t=4$ -- 6.
This makes the estimation of the continuum value of $\sigma$ difficult.

A similar violation of scaling was also seen in the measurement of
the latent heat from the same data by QCDPAX
collaboration~\cite{QCDPAX92}.  Because the pressure $P$ is continuous
across the phase transition, the latent heat $\latent$ can be calculated from
the discontinuity of $(\eps-3P)/T^4$.
On an Euclidean lattice this is proportional to the discontinuity in
the average plaquette $<U_\Box>$:
\be
\Delta (\eps-3P)/T^4 = -72 N_t^4\tilde\beta(g)/g^3 \Delta <U_\Box>\,
\ee
where $\tilde\beta(g) = -a\partial g/\partial a$ is the
renormalization group beta function.  $\tilde\beta(g)$ can be
evaluated with the 1(2)-loop perturbative analysis, or, since the
violation of asymptotic scaling at these $\beta$'s is well
established, with non-perturbative MCRG methods~\cite{Hoek90}.  The
quantity $\eps-3P$ is preferred in lattice calculations over directly
measuring $\eps$ or $\eps+P$, because the non-perturbative corrections
to the latter quantities are not known.  For $N_t=2$, we use the
1-loop beta function and the plaquette gap measured by Alves
\etal\cite{Alves92}, and for $N_t=4,6$ we use the results of QCDPAX
collaboration~\cite{QCDPAX92}, evaluated
with the 1-loop beta
function and with
the non-perturbative beta function.  The values are collected in
Table~\ref{table2}.

\begin{table} \center
\begin{tabular}{cccccc} \hline
$N_t$ & $\sigma/T^3$ & 
$\latent_\rmi{1-loop}/T^4$ &$\latent_\rmi{MCRG}/T^4$ 
&$\alpha_\rmi{1-loop} $    &$\alpha_\rmi{MCRG}$ \\ \hline 
2 & 0.092(4)\n\n& 2.48(5)\n\n&  --     & 0.046(3)\n\n  &   --       \\
4 & \SigRotF    & 4.062(85)  &2.44(24) & 0.0050(6)\n   & 0.0084(13) \\
6 & \SigRotS    & 2.395(63)  &1.80(18) & 0.0055(13)    & 0.0073(18) \\\hline
\end{tabular}
\caption[1]{The interface tension $\sigma$, 
the latent heat $\latent = \Delta(\eps-3P)$, and the scale factor
$\alpha = [16\pi\sigma^3/(3\latent^2 T)]^{1/2}$ for different $N_t$'s.
$\latent$ and $\alpha$ are evaluated with the 1-loop and MCRG beta
functions.\la{table2}}
\end{table}

{F}rom $\latent$ and $\sigma$ we can now calculate the ratio $\alpha^2
= 16\pi\sigma^3/(3\latent^2 T)$, which is also shown in
Table~\ref{table2}.  We observe that when $N_t = 4$ -- 6, the scaling
violations seem to be absent in $\alpha$, at least within the accuracy
reached in this work.  This holds both for the 1-loop and MCRG
corrected values.  On the other hand, the $N_t=2$ value is completely
off the mark when compared with the higher $N_t$ results.  

Inserting
the $N_t=6$ value of $\alpha_\rmi{MCRG}$ to the
\eqs\nr{bubble1} and \nr{bubble2}, 
we find that the degree of supercooling at the
deconfinement phase transition in the early universe and the average
distance between the nucleation centers ($\sim$ the scale of the
inhomogeneities generated during the transition) are given by
\be
 (T_c - T_{\rm pt})/T_c = 5.6(1.4) \times 10^{-4}\,, 
\ee
and
\be
	\h\h d = 22(5)\mbox{\,mm}\,,
\ee
respectively.
Another interesting quantity is the ratio $\sigma T/\latent$, which is
relevant when one calculates the hydrodynamical properties of the
propagating phase transition front \cite{Kajantie92}.  These values
are shown in Table~\ref{table3}.  For $N_t=4$ -- 6, the ratio
$\sigma T/\latent$ also appears to be scaling within the statistical
errors.

\begin{table} \center
\begin{tabular}{ccc} \hline
$N_t$ &$(\sigma T/\latent)_\rmi{1-loop} $  
      &$(\sigma T/\latent)_\rmi{MCRG}$ \\ \hline 
2 & 0.037(2)\n\n &   --       \\
4 & 0.0072(6)\n  & 0.0120(15) \\ 
6 & 0.0091(14)   & 0.0121(22) \\ \hline
\end{tabular}
\caption[1]{The ratio $\sigma T/\latent$, evaluated both with the
1-loop and MCRG improved beta functions.\la{table3}}
\end{table}

\section{Conclusions}
\label{conclusions}

We have measured the interface tension $\sigma$ between the confining
and deconfining phases of SU(3) lattice gauge theory by the
histogram method using the data from the high-statistics simulations on
$N_t=4$ and 6 lattices by QCDPAX collaboration~\cite{QCDPAX92}.
The main advantage of the histogram
method is that it offers a direct way to measure numerically the
interface tension from order parameter distributions.
Hence, the results of ordinary finite temperature simulations near
the transition point can be used.
We also made a careful analysis of finite size corrections to
the interface tension in order to 
estimate the infinite volume values.  
We found 
our ansatz for finite size corrections is quite reasonable
in the sense that our data are well fitted to the formula with
quite acceptable $\chi^2$ values.

The results, $\sigma/T_c^3 = \SigRotF$ and \SigRotS\ for $N_t=4$
and 6, respectively, exhibit a scaling violation qualitatively similar 
to the latent heat $\latent$~\cite{QCDPAX92}.  
Since both the latent heat $\latent$ and the interface tension $\sigma$
characterize the strength of the first order transition, it may not
be surprising that we observe similar scaling violations.
Thus, it is plausible that when one takes a suitable combination
of $\latent$ and $\sigma$, scaling becomes much better.
Indeed, we observed that the
physically relevant combinations $\alpha^2 \propto
\sigma^{3}T/\latent^2$ and $\sigma/\latent T$ scale 
within the statistical errors.
Clearly, scaling for the dimensionless combinations and the lack
of scaling for $\latent$ and $\sigma$ themselves are not satisfactory.  
However,
we can use the combinations that do scale to predict the continuum physics.

Nevertheless, one has to be very cautious when applying these numbers
to cosmology.  We have neglected the effect of quarks, which
may change the character of the transition completely. 
However, we believe that our results offer some intuition about the
expected physical scales relevant for the QCD phase transition.

\subsubsection*{Acknowledgments}

We are grateful to Y.~Aoki,
A.~Irb\"ack, K.~Kajantie and A.~Ukawa for useful discussions.  This work
was partly supported by the Grants-in-Aid of Ministry of Education,
Science and Culture of Japan (No.~62060001 and No.~02402003) and by
the U.~S.~Dept. of Energy grant No.~DE-FG02-8SER40213.

\pagebreak

\pagebreak

\subsection*{Figure captions}

\begin{figure}[h]
\caption[1]{The probability distributions of the rotated Polyakov loop,
normalized to $p_{\max}=1$:
(a) for $N_t=4$ with two different volumes; (b) for $N_t=6$ with three
different volumes. 
\label{fighisto}}
\end{figure}

\vfill\mbox{}

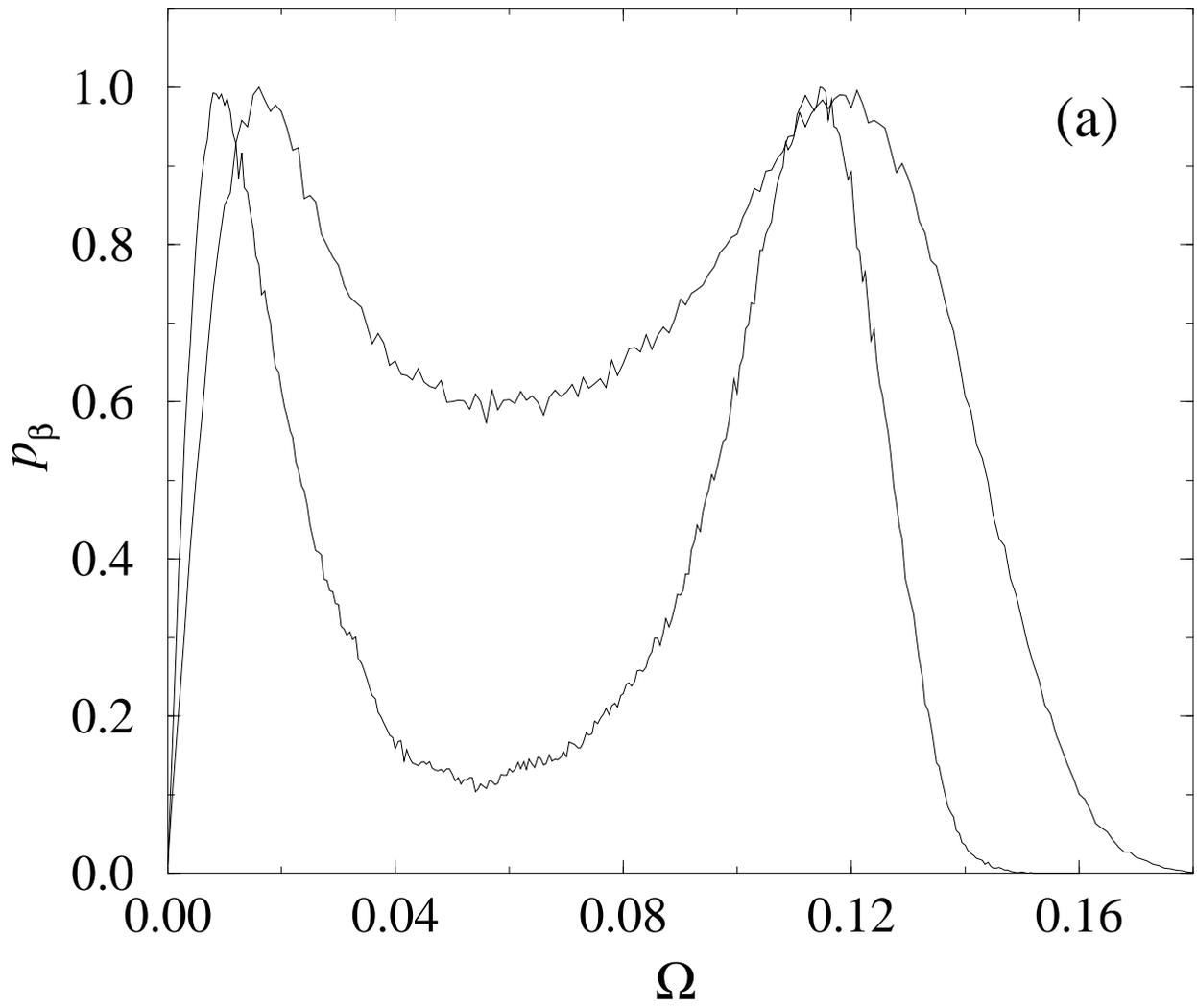
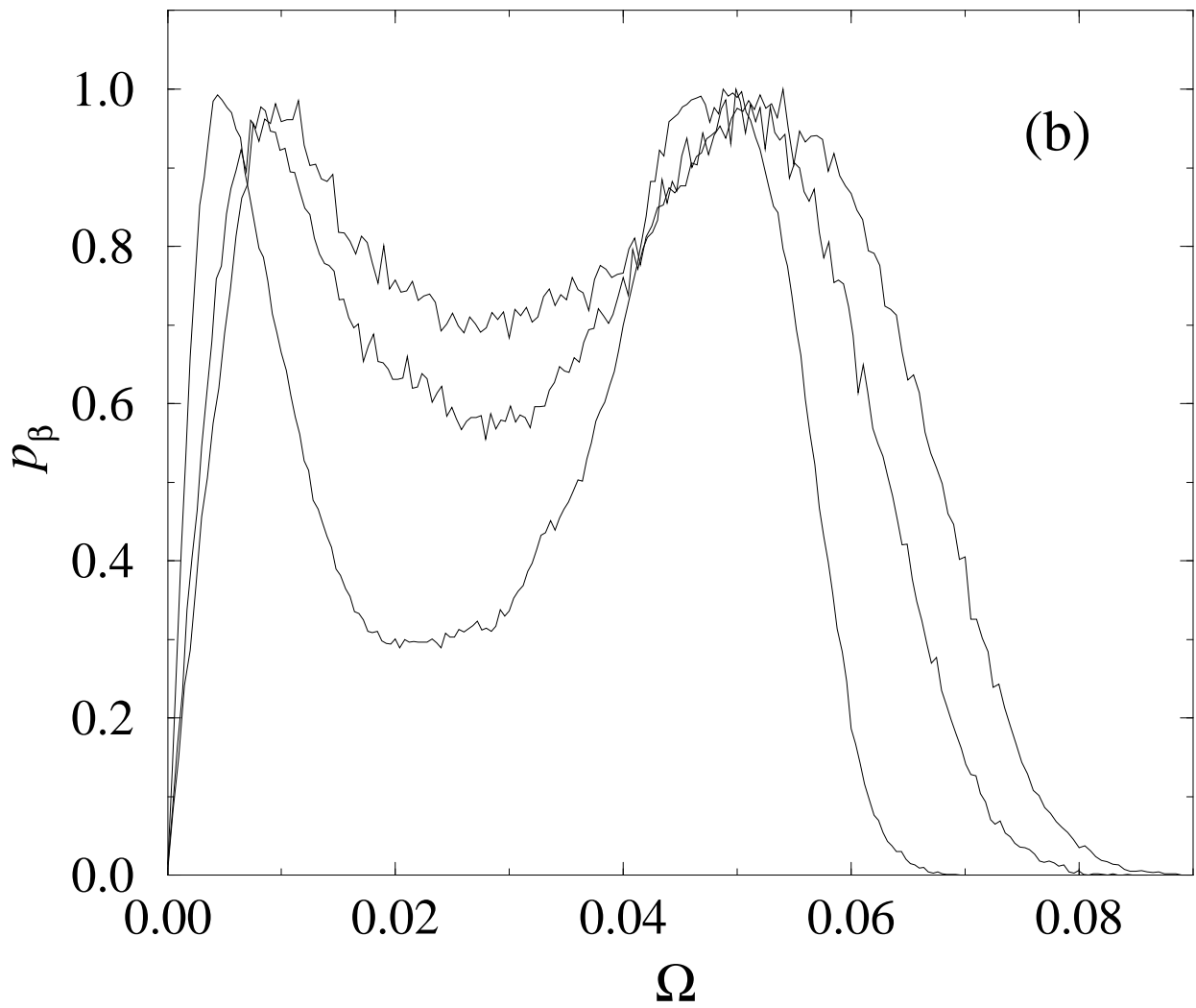
\end{document}